\documentclass[useAMS,usenatbib]{mn2e}
\usepackage{graphicx}
\usepackage{epsfig}
\usepackage{natbib}  
\usepackage{epstopdf}
\newcommand\aj{{AJ}}%
\newcommand\araa{{ARA\&A}}%
\newcommand\apj{{ApJ}}%
\newcommand\apjl{{ApJ}}%
\newcommand\apjs{{ApJS}}%
\newcommand\aap{{A\&A}}%
\newcommand\aaps{{A\&AS}}%
\newcommand\mnras{{MNRAS}}%
\newcommand\pasp{{PASP}}%
\newcommand\nat{{Nature}}%
\def\simgt{\lower.5ex\hbox{$\; \buildrel > \over \sim \;$}}
\def\simlt{\lower.5ex\hbox{$\; \buildrel < \over \sim \;$}}

\newcommand\rt{{\rm r}$_t$}
\newcommand\rh{{\rm r}$_h$}
\newcommand\rj{{\rm r}$_J$} 
 
\newcommand\teff{T$_{\rm eff}$}

\newcommand\Teff{T$_{\rm eff}$}

\newcommand{\msun}{\ensuremath{\, {M}_\odot}}
\newcommand{\Msun}{\ensuremath{\, {M}_\odot}}
\newcommand{\ocen}{$\omega$~Cen}

\title[In search of single-population globular clusters]{In search of massive single--population 
Globular Clusters
}
\author[V. Caloi \& F. D'Antona]
{Vittoria Caloi$^{1}$ and Francesca D'Antona$^{2}$
\thanks{E-mail: dantona@oa-roma.inaf.it (FD); vittoria.caloi@iasf-roma.inaf.it (VC)}
\\
$^{1}$ INAF, IASF--Roma, via Fosso del Cavaliere 100, I-00133 Roma, Italy\\
$^{2}$ INAF, Osservatorio Astronomico di Roma, Via Frascati 33, 
I-00040 Monteporzio Catone (Roma), Italy.\\
}
\begin{document}

\date{Accepted . Received ; in original form }

\pagerange{\pageref{firstpage}--\pageref{lastpage}} \pubyear{2006}

\maketitle

\label{firstpage}

\begin{abstract}
   The vast majority of globular clusters so far examined shows the chemical signatures of hosting (at least) two stellar populations. According to recent ideas, this feature requires a two--step process, in which the nuclearly processed matter from a ``first generation" (FG) of stars gives birth to a ``second generation" (SG) bearing the fingerprint of a fully CNO--cycled matter. Since, as observed, the present population of most globular clusters is made up largely of SG stars, a substantial fraction of the FG ($\simgt 90$\%) must be lost.  Nevertheless, two types of clusters dominated by a simple stellar population (FG clusters) should exist: either clusters initally too small to be able to retain a cooling flow and form a second generation (FG--only clusters), or massive clusters that could retain the CNO processed ejecta and form a SG, but were unable to lose a significant fraction of their FG (mainly--FG clusters). Identification of mainly--FG clusters may provide an estimate of the fraction of the initial mass involved in the formation of the SG.

 We attempt a first classification of FG clusters, based on the morphology of their horizontal branches (HBs), as displayed in the published catalogues of photometric data for 106 clusters. We select, as FG candidates, the clusters in which the HB can be reproduced by the evolution of an almost unique mass. We find that less than 20\% of clusters with [Fe/H]$<$--0.8 appear to be FG, but only $\sim$10\% probably had a mass sufficient to form  at all an SG.  This small percentage confirms on a wider database the spectroscopic result that the SG is a dominant constituent of today's clusters, suggesting that its formation is an ingredient necessary for the survival of globular clusters during their dynamical evolution in the Galactic tidal field.

In more detail we show that Pal~3 turns out to be a good example of FG--only cluster. Instead, HB simulations and space distribution of its components,  indicate that M~53 is a  ``mainly--FG" cluster, that evolved in dynamical isolation and developed a small SG in its core thanks to its large mass. Mainly--FG candidates may be also NGC~5634, NGC~5694 and NGC~6101. In contrast, NGC~2419 contains $>$30\% of SG stars, and its present dynamical status bears less information on its formation process than the analysis of the chemical abundances of its stars and of its HB morphology. 
\end{abstract}


\section{Introduction}
\label{sec:intro}

 A general finding of recent years is that all globular clusters (GCs) so far spectroscopically examined contain multiple stellar populations. This is significatively shown in the recent analysis of \cite{carretta2009a} of about two thousand stars in 19 GCs, showing that all these clusters display the sodium -- oxygen anticorrelation, signature of the presence of a population of stars sodium richer and oxygen poorer than the halo stars of the same
metallicity. The Na--O anti--correlation is typical of GCs, whose constituent stars belong to two or more stellar populations differing in the abundances of the elements produced by the hot CNO cycle and by other proton--
capture reactions on light nuclei.  In fact, these chemical signatures are present also in turn--off stars and among the subgiants \citep[e.g.][]{gratton2001,briley2002, briley2004}, so they can not be imputed to ``in situ" mixing in the stars, but must be due to some process of self--enrichment occurring at the first stages of the cluster life.
Photometric evidences for the presence of multiple populations are also numerous, and sometimes suggestive of star formation occurring in separate successive bursts. The photometric signatures of different populations can be imputed in part to helium differences, inferred from the morphology of the horizontal branches (HB) \citep{dantona2002,dc2004,lee2005}, or from the presence of multiple main sequences, \citep{norris2004, piotto2005, piotto2007}. 

  In view of all this, at present the formation of globular clusters is   considered a two--step process lasting no longer than $\sim$100 Myr, during which the nuclearly processed matter from a ``first generation" (FG) of stars gives birth, in the   cluster innermost regions, to a ``second generation" (SG) of stars with  the characteristic signature of a distribution of element abundances fully CNO--cycled.

  In this regard, a major problem remains and is rarely faced: the spectroscopic information shows that, in the clusters so far examined, the percentage of stars of the SG is generally $\sim$50--80\%  \citep{carretta2009a}, as also results by interpreting the HB morphologies in terms of helium enrichment  \citep{dantonacaloi2008}. This large percentage can  not be the result of chemical evolution within a ``closed box", simply  because the processed matter available from the more massive stars is always a  small percentage of the FG mass\footnote{In the case of massive asymptotic giant branch polluters, the mass consistent with the chemical anomalies (including both processed ejecta and diluting gas) goes from $\sim$8 to 12\% of the initial cluster mass, depending on the IMF, see e.g. Vesperini et al. 2010.}. Anomalous initial mass functions of the FG may help, but  they pose further problems for the dynamical survival of clusters; instead, it seems necessary that the matter forming the SG stars is collected from a much larger stellar ensemble. In  other words, the cluster has managed to lose most of its FG stars, and is now the ``small" remnant of the evolution of a much more massive  --and maybe also of a much bigger-- stellar system.

  So the first proposal on the subject, by Bekki \& Norris (2006), assumes  that all GCs are formed within dwarf galaxies, that are now dispersed. This kind of  formation is generally accepted to have occurred in the most massive clusters, such as \ocen , M22, and in M54, the cluster belonging to the Sagittarius dwarf galaxy  \citep{ibata1994,bellazzini2008}.  These clusters show also metallicity spreads  \citep[e.g., among others][]{norrisdacosta1995, marino2009, carrettam542010}  that are not present in most of the other clusters. More difficult it is to believe that  {\it all} GCs have formed within dwarf galaxies. We point out, e.g., that old massive clusters in the Magellanic Clouds also show the Na--O anticorrelation \citep{mucciarelli2009}.
  
A different proposal to explain the presence of multiple stellar populations in Galactic GCs, with the observed number fractions of FG and SG stars, was advanced by \cite{dercole2008}. The model is based on the loss of most of the FG stars at early phases, during the formation of the SG stars in a cooling flow at the cluster center. The idea in support of this model is that the mass loss from the explosion of supernovae type II (SN~II), and the associated loss of the remnant gas from which the FG stars had formed, occurring just previous to the formation of the SG, produce an expansion of the cluster, leading to a loss of a significant fraction of FG stars. \cite{dercole2008} show that 90\% or more of the initial FG mass may be lost, so that the SG may become an important fraction of the total stellar population, even overcoming the FG. However, if the cluster total size was strongly under filling its tidal radius, or the cluster evolved in isolation, the FG can not be lost and the cluster dynamically survives to the SN~II epoch, maybe leaving  a looser cluster structure as fingerprint of the mass loss from the SN~II explosions \citep[e.g.][]{bastian2008,vesperini2009}. In this case, the SG formation remains {\it a small perturbation in the cluster history}, and can not represent more than a  few percent of the total (mainly--FG cluster).  

Of course, true ``FG--only" clusters should exist: those in which a SG could not form because their inital mass was too low to allow for the formation of a cooling flow \citep{dercole2008, vesperini2010, bekki2011}. These clusters will generally survive only if they do not interact strongly with the Galactic gravitational field, otherwhise the expansion due to the SN~II mass loss will lead to the cluster disruption.  The fact that most GCs so far examined host a large fraction of SG stars seems to imply that the very formation of a SG ---that occurs after the SN~II epoch and is not subject to the cluster expansion--- and its dynamical interaction with the FG stars allowed massive clusters to survive in a tidally limited environment \citep{dvmessenger2008}, although they lose more than 90\% of the initial mass. 

A dynamical identification of these two important classes of clusters (FG-only and mainly-FG) is particularly complicated as it would require a reliable reconstruction of the individual cluster dynamical histories and initial structural properties. For example, NGC 2419 is strongly underfilling its Jacobi radius (\rj), the tidal radius of the cluster due to the galactic field, computed in the plain Roche approximation \citep[e.g.][]{portegies2010}, defined by its present mass and  Galactocentric distance; on the basis of its current structural properties and  position in the Galaxy, one might naively consider NGC~2419 to be a good candidate of a ``mainly-FG" cluster. However as discussed in  \cite{cohen2010} and \cite{dicriscienzo2011}, its eccentric orbit \citep{casetti2009} and/or, possibly,  a different galactic environment in its early stages of evolution must  have caused a large mass loss, and this cluster appears to contain a significant fraction of SG stars.

We point out that the formation of the SG  subsystem is likely to have a significant impact on the cluster structural properties. While young clusters in nearby galaxies form compact, with a half--mass radius \rh$<$1~pc \citep{lada2003}, they expand thanks to mass loss by stellar winds and SN~II explosions  in the first 30Myr of life \citep[e.g.][]{bastian2008}, and this expansion may be enhanced if the cluster is initially mass segregated \citep{vesperini2009}. 
If the SG forms in a cooling flow, this leads naturally again to an even more compact stellar distribution \citep{dercole2008}, with a small \rh \footnote{As shown by \cite{vesperini2011}, a further indication that the SG is more centrally concentrated than the FG lies in the lack of barium stars in the SG of several clusters \citep{dorazi2010}.}. Otherwise, if the SG does not form, the cluster maintains  the larger  \rh\ acquired in the first expansion phase, so a cluster may be tidally filling \citep[e.g. \rh/\rj$>$0.1,][]{baumgardt2010} simply because it has not developed an SG.
\begin{table*}
\caption{Candidates FG clusters, plus  NGC 2419}             
\label{tab}      
\centering          
\begin{tabular}{c c c c r r r r r r c}     
\hline\hline       
Name        & [Fe/H]  & M$_{V}$  & log M$_c$  & d$_\odot$(kpc)  & d$_{gc}$(kpc) & r$_c$(pc) & r$_h$(pc) & r$_J$ &  RR ~Lyr &  HB \\ 
\hline            
NGC 4372 & --2.09  & --7.79 &  5.34  &5.8     & 7.1    & 3.92  & 8.75  &  78.8    &    0 (A) &  B, d (a) \\
NGC 5024 (M 53) & --1.99  & --8.70 &  5.71 & 17.8 & 18.3  &   2.48& 7.66 &  197.0 & 59 (n) &  B, p (n)  \\
NGC 5634 & --1.88  & --7.69 &  5.31  &25.2  & 21.2  & 2.05  & 5.27   &  160.0 &  20 (B) &  B, p (b) \\
NGC 5694 & --1.86  & --7.81 &  5.36  &  34.7& 29.1  & 0.80 & 4.43    & 205.5      &    0  (A)&  B, p (c) \\
NGC 5897 & --1.80  & --7.21 &  5.12   &12.4 & 7.3    & 9.40  & 10.12  &   67.8  &  11 (C)  &  B, p? (d,e,f) \\
NGC 6101 & --1.82  & --6.91 &  5.00   &15.3 & 11.1  & 6.81  & 10.12   &   81.7 &    0  (A)&  B, p (g) \\
NGC 6139 & --1.68  & --8.36 &  5.58   &10.1 & 3.6    & 0.54  & 3.20    &   60.2  &    4 (A) &  B, d (c)  \\
NGC 6235 & --1.40  & --6.44 &  4.82  & 11.4 & 4.1    & 1.59  & 3.70   & 36.8  &  3 (A) &  B, d (c,h)\\
NGC 6652 & --0.96  & --6.66 &  4.91   &10.1 & 2.7    & 0.28   & 2.54    &   29.7 &  0  (A) & R (c) \\
NGC 6717 & --1.29  & --5.66 &  4.50   &7.1   & 2.4    & 0.22   & 1.86    &   20.1 &  1 (A)  & B, d (a,c) \\
Arp 2         & --1.76  & --5.29 &  4.35   &28.6  & 21.4 & 17.60  & 21.13  &   76.9 &  9 (B)  & B, d (i) \\
Terzan 8    & --2.00  & --5.05 &  4.26   & 26.0 & 19.1 & 10.06  & 10.06  & 66.7  &  3 (B) & B, p? (j) \\
\hline 
AM1         & --1.80  & --4.71 &  4.12 & 121.9    & 123.2    &7.09   & 23.64   &  207.0   &  0 (A)&  R (k)  \\
Eridanus  & --1.46  & --5.14 &  4.29  & 90.2    & 95.2       &8.74   & 14.00    & 198.6   &  0 (A)&  R (l)  \\
Palomar 3 & --1.66  & --5.70 &  4.51      & 92.7 & 95.9    & 17.25  & 23.7    &  236.4   &  7 (j)  &  R (j) \\
Pal 4        & --1.48  & --6.02 &  4.64  & 109.2   & 111.8     &23.29 & 22.87   &  289.2   &  0 (A)&  R (l)\\
Pal 14      & --1.52  & --4.73 &  4.13   &  73.9  & 69.0       &26.94 & 32.96    &  141.7   &  0 (A)&  R (k)\\
\hline
NGC 2419  & --2.12  & --9.58 &  6.06  & 84.2 & 91.5    & 11.40    & 23.78  &   753.0    & 75 (D) & B, p+bh (m)\\
\hline 

\end{tabular}
\\
{$^{(a)}$ Brocato et al. 1996}; ~{$^{(b)}$ Bellazzini et al. 2002}; {$^{(c)}$ Piotto 2002}; {$^{(d)}$ Ferraro et al. 1992} ; \\  {$^{(e)}$} Sarajedini 1992 ; {$^{(f)}$} Testa et al. 2001 ; {$^{(g)}$} Marconi et al. 2001;  {$^{(h)}$ Howland et al. 2003};  {$^{(i)}$ Buonanno et al. 1995};  {$^{(j)}$ Montegriffo et al. 1998};  {$^{(j)}$ Hilker 2006};  {$^{(j)}$ Stetson et al. 1999}; {$^{(m)}$ Di Criscienzo 2011b; {$^{(n)}$ Rey et al. 1998}.}   \\
{$^{(A)}$ Clement et al. 2001}; {$^{(B)}$ Salinas et al. 2005}; {$^{(C)}$ Clement \& Rowe 2001}; {$^{(D)}$ DiCriscienzo 2011a}
\end{table*}

Considering the possible ambiguities in the identification of FG--only and mainly--FG clusters from dynamical information, we have decided to resort to a photometric criterion, that of the evolutionary status of their HBs. In this work we examine the existing astronomical literature to identify clusters that we expect to be either FG--only (low initial mass) or mainly--FG (high initial mass) clusters. If we can identify mainly--FG clusters, and discover the presence of a small percentage of SG stars in them, this will help constraining the model for the formation of multiple generations in GCs.

\section{Selection of FG--only clusters}
\label{two}
One easy way to recognize the presence of a second generation in a GC is  to consider the morphology of the HB \citep[e.g.][]{dantona2002, gratton2010}.  Historically, the dispersion in mass along the HB was imputed to a dispersion in mass loss \citep{rood1973}, but recent developments have shown that it may be in large part due to differences in helium content, that appear together with the chemical signatures of the second generation \citep[][and references therein]{marino2011}. As discussed in \cite{dantonacaloi2008}, bimodal HBs, blue tails, gaps in the stars distribution and, in general, HBs that extend from the red to the faint blue are the most clear candidates for the presence of multiple stellar generations. As for less extreme morphologies, the case of M~3 was examined in detail by \cite{caloi2008}.
While an appropriate dispersion in mass loss \citep[$\sigma \sim 0.02$\msun][]{catelan2001} allows to reproduce the ratios in number among the HB components (red, variable, blue), the detailed color distribution along the HB and the RR~Lyraes peaked period distribution can not be reproduced by a ``normal" population with mass spread. On the contrary, a generation of stars with normal helium, with a very small mass dispersion ($\sigma \simlt 0.003$\msun) can account very well for the red and variable HB members, while the blue region is populated by a second stars generation with variable helium. The very good fit obtained in these conditions for the number vs. period distribution of RR Lyr stars, otherwhise unattainable \citep{catelan2004, castellani2005} strongly supports this interpretation. 
\begin{figure*}
\begin{center}
\includegraphics[width=5cm]{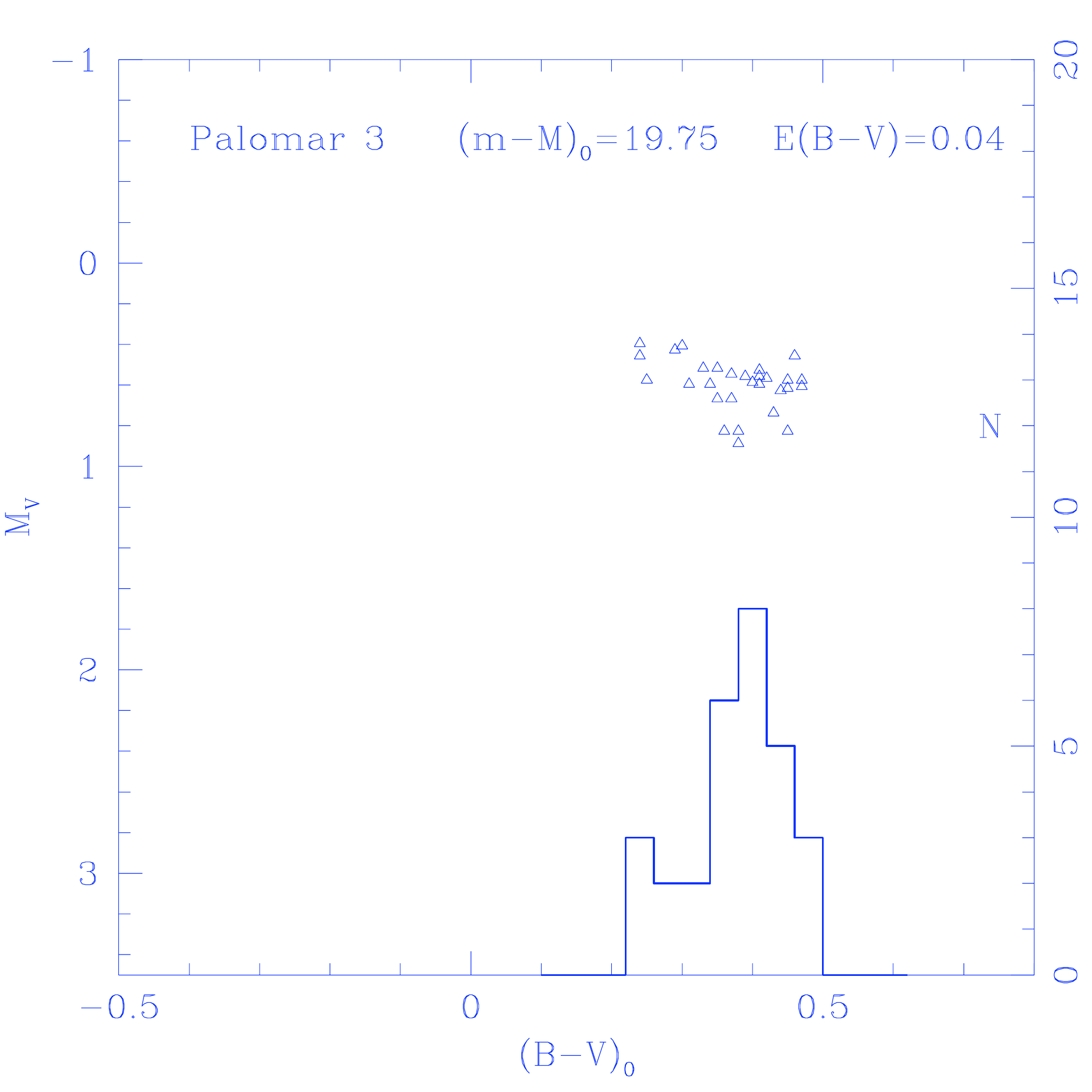}
\includegraphics[width=5cm]{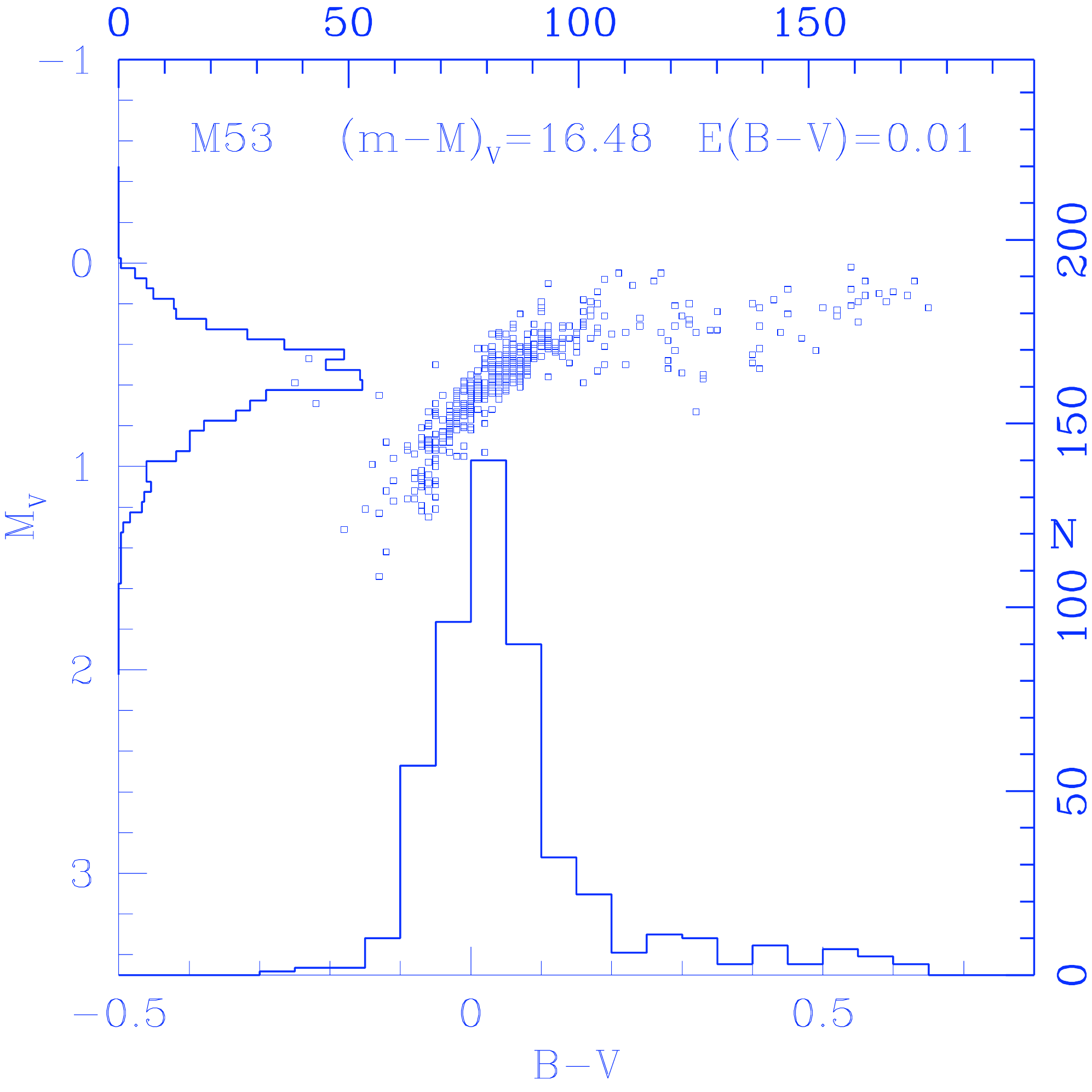}
\includegraphics[width=5cm]{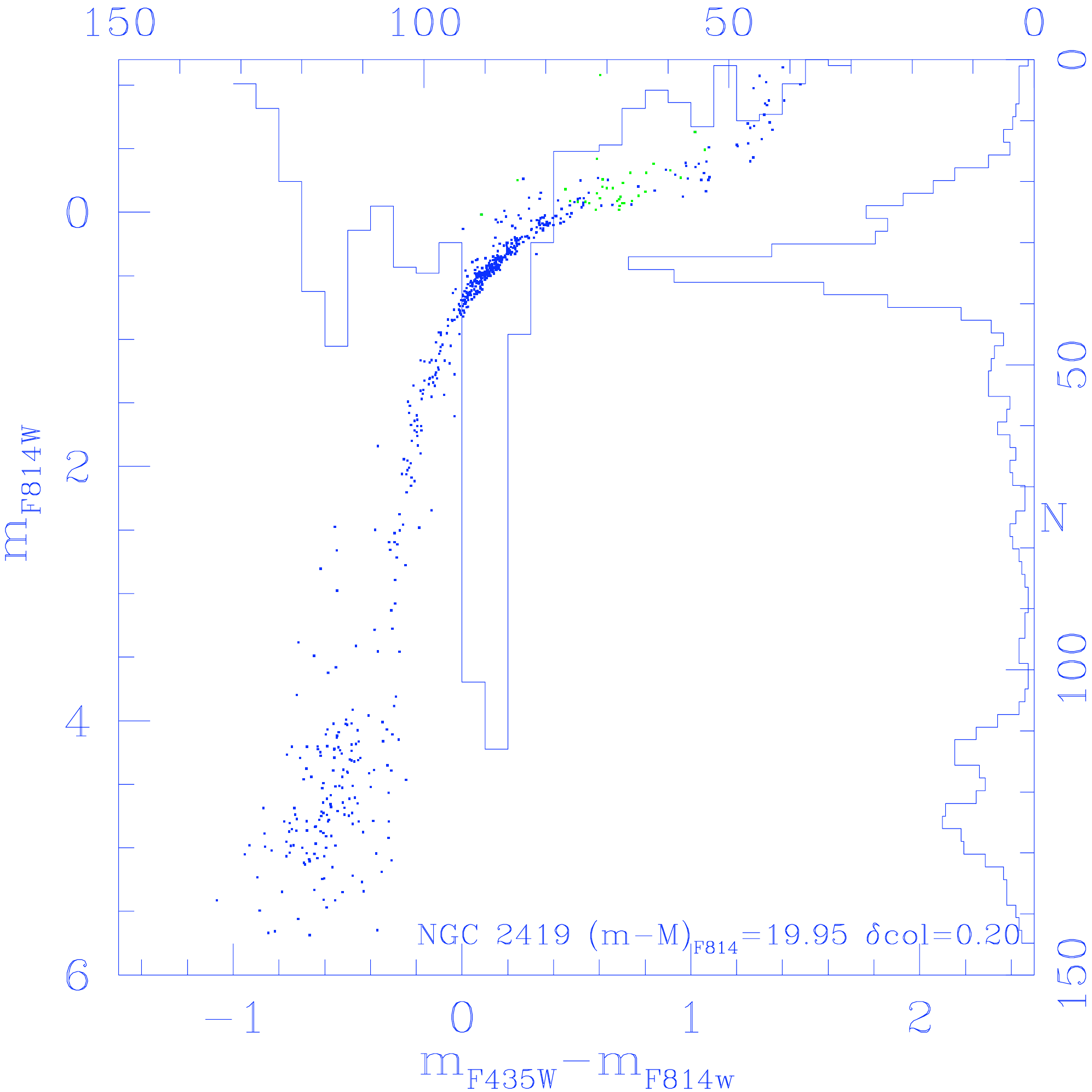}
\caption{HB stellar distributions for three very different cases. Left panel: Pal~3 from Hilker 2006; central panel: M~53 data from Rey et al. 1998; right panel: NGC~2419 from Di Criscienzo et al. 2011. The histograms represent the number counts of HB stars as function of the colour (all panels) and of the magnitude (central and right panel).}
\label{f1}
\end{center}
\end{figure*}

In this framework, the single population GCs should be identified by an HB that can be reproduced by {\it the evolution of an almost unique mass}. We adopt the presence of a ``short" HB as a first indication of an FG cluster, and then consider its dynamical status. To single out the candidates we examined the color magnitude (CM) diagrams in the databases by \cite{rosenberg2000a,rosenberg2000b} and by \cite{piotto2002}, covering a total of  96 GCs. Besides, we took into account  five other clusters (NGC 5686, NGC 6749, NGC 7492, Arp2, Terzan 8) for a total of 101 CM diagrams present in the current literature. Clusters with bad CMs, insufficient for our purposes (e.g., CM diagrams of clusters in the bulge or projected onto it), have not been considered. Besides, we exclude those clusters with [Fe/H]$>$--0.8 that show an exclusively red HB. In these cases, even if an SG is present, it is not easily identifiable with simple photometric criteria \citep[see, e.g., the case of 47 Tuc discussed in][]{dicriscienzo2010}. 

After this first screening, we are left with 86 CM diagrams, among which FG clusters must be identified. Only one cluster is left with a red HB (NGC~6652). For the others, whose HB is blue, we select those in which the extension in V magnitude of the HB does not exceed 1~mag. We further exclude the clusters (with a short blue HB) for which the O--Na anticorrelation has been observed, namely: M30, NGC 6397 \citep{carretta2009a} and NGC 7492 \citep{cohen2005a}. When available, we checked the results on more than one CM diagram. These checks allowed to eliminate clusters that in a first moment looked like FG. A clear example is given by NGC 6535, that looks like hosting a blue HB of about 1 visual magnitude range in the data by \cite{testa2001}. A look at the photometry by Sarajedini (1994) shows that the HB extends for $\sim$1.8 visual magnitudes, and, further, that it contains two very faint extreme HB stars. On this basis, this cluster was excluded. 

We consider, in addition, a group of small mass (1.5--4.5 $\times 10^4$\msun), far away clusters (galactocentric distance d$_{gc} \simgt$70Kpc): AM1, Eridanus, Pal 3, Pal 4, Pal 14, apparently 1--2~Gyr younger than M 3, but of similar metallicity. Their HBs are short, exclusively red, except for Pal 3, in which 7 RR~Lyr variables are found.  They all appear as good candidates for FG--only clusters. 

Out of the 86 clusters we selected 12 clusters as FG (14\%). To this figure we may add the 5 clusters listed above. This constitutes $\sim$19\% of the galactic GCs for which we have reasonably good CM diagrams. This estimate may be a lower limit (as we have excluded the red HB, metal rich clusters), but a detailed exam of the sample may also show that it is an upper limit. Notice in fact that three clusters with the same photometric characteristics have been excluded on the basis of spectroscopy, that is not yet available for the clusters in the list.  

Data for the selected clusters are presented  in Table 1, where we also list NGC~2419, to be discussed later.  [Fe/H], absolute visual magnitude M$_{\rm v}$\ and the distances from the Sun (d$_\odot$) and from the Galactic center (d$_{gc}$) are taken from Harris (2003). The mass is computed by assuming a mass to visual luminosity ratio of two. The half mass radius is either directly taken from \cite{baumgardt2010}, or computed according to their prescription (assuming that the true half mass radius is 1.33 times the projected half mass radius given by Harris 2003). The Jacobi radius is computed according to \cite{baumgardt2010} prescription. In column 10 we list the number of known RR~Lyr stars, if any, and in column 11 we indicate the predominant color of the HB (generally B=blue, only one R=red cluster is present) and whether the star distribution appears  peaked (p, like in M~53) or more distributed (d). The presence of RR~Lyraes is a further indication of the extension in color of the HB, although the fact that they are generally very few indicates that the RR~Lyr gap is likely traversed by stars evolving out of the ZAHB (see later).

Several of the 17 selected clusters have a small (present) mass. It is possible that also in the past this mass was small enough that the clusters could not form the SG stars. Recent hydrodynamic 1D computations by \cite{vesperini2010}, covering a wide range of cluster structural parameters, show that all the ejecta that may form SG stars are retained above $\sim 10^6$\msun\ of {\sl initial} cluster mass, while the retention is  very limited for all models of inital mass $\simlt 10^5$\msun. Taking into account that clusters in Table~1 may have lost mass, due to two--body relaxation processes and tidal shocks, we may adopt a conservative limit of log~(M$_c$/\msun$) <$4.8 as a formal dividing line below which clusters do not form SG stars (FG--only). This choice selects 8 clusters in Table~1, seven of which are also in the list  of  tidally filling clusters at d$_{gc} >8$~kpc by \cite{baumgardt2010}. The other clusters in common between ours and \cite{baumgardt2010} list are the compact clusters (\rh/\rj$<$0.05) M~53, NGC~5694  and NGC~5634.  We suggest than that the small mass clusters have a larger \rh\ because {\it they did not form the SG}, as outlined in the Introduction.

In our list, we are left  with 9 clusters ($\sim$10\%) that may have initially developed an SG. 
This small fraction is consistent with, and extends,  the spectroscopic result by \cite{carretta2009a, carretta2010b}, who find in their whole 19 clusters sample a predominant ($>$50\%) SG. We must conclude that very few of the clusters that may develop the SG do not lose a high fraction of their initial mass, or even that only a few clusters that do not form a SG {\it survive} to the dynamical interaction with the Galactic tidal field. The formation of a central compact SG system appears to be  a key ingredient for the survival of a cluster to the first phases of cluster evolution  \citep[][]{dvmessenger2008, dercole2008} .

  \section{The cases of Palomar 3 and M 53 }
  
 We examine in detail two clusters: Pal 3 (an example of FG--only cluster) and M~53 (for which we pose the case of a mainly--FG cluster).  For comparison, we also discuss the case of NGC 2419: although it is now evolving fully inside its Jacobi radius, the cluster contains a substantial extreme SG, implying a peculiar dynamical history. The HBs of these three clusters are shown in Fig.~\ref{f1}, together with the histograms of the color and magnitude distribution of HB stars. It is straightforward to appreciate that Pal~3 HB is extremely short and that the HB of M~53 shows a very peaked distribution in color, with a tail of redder stars. The strikingly different HB  of NGC~2419 shows a strongly peaked color and magnitude distributions for its more luminous HB component, as in M~53, but has also a long tail of bluer stars, ending with a blue hook.

\subsection{Simulations}

The HB simulations for Pal~3 and M~53 are based on the models published in \cite{dantona2002} for metallicity Z=2$\times 10^{-4}$ and Z=10$^{-3}$, with solar--scaled $\alpha$--elements abundances. Helium contents Y=0.24 and Y=0.28 are considered. The simulations for NGC~2419 are based on the models presented in \cite{dicriscienzo2011}, for a mixture with [Fe/H]=--2.4 and [$\alpha$/Fe]=0.2. Helium contents of Y=0.24, 0.28 and 0.42 have been considered. A detailed description of models is given in the quoted papers.

Synthetic models for the HB are computed according to the recipes described in \cite{dantonacaloi2008}. We adopt the appropriate relation between the mass of the evolving giant $M_{RG}$ and the age, as function of helium content and metallicity. The mass on the HB is then: 
\begin{equation} 
M_{HB} =  M_{RG}(Y,Z) - \Delta M 
\end{equation} 
$\Delta M$\ is the mass lost during the RG phase. We assume that $\Delta M$\  has a Gaussian dispersion $\sigma$\ around an average value $\Delta M_0$\  and that both $\Delta M_0$\  and $\sigma$\ are parameters to be determined and {\it in principle} do not depend on Y.  Once chosen Z and Y, the  \teff\  location of an HB mass is fixed. Consequently, different ages can be adopted, provided that the mass loss is consistently adjusted. 
For HBs extending into the variability region, the RR~Lyraes are identified as those stars that, in the simulation, belong to the \Teff\ interval $3.795 < \log T_{\rm eff} < 3.86$. Their periods are computed according to the pulsation equation (1) by Di Criscienzo, Marconi \& Caputo (2004).

  \subsection{Palomar~3}
     Palomar~3 is a remote cluster at about 96~kpc from the Galactic center and an estimated orbital minimum distance from it of $\sim$ 82.5 kpc \citep{dinescu1999}. Since its proper motion is uncertain, it is considered possible that it may not be bound to the Galaxy and that it may be falling onto it for the first time. Pal~3 is one of the most extended clusters with a half--light radius of about 24~pc and a truncation radius of about 130~pc. It is faint, with $M_{V}$ $\sim$ --5.7, and its destruction time is estimated at about 20 Hubble times \citep{gnedin1997}.

     Two colour--magnitude (CM) diagrams of Pal~3 are available \citep{stetson1999, hilker2006}. This tiny cluster presents a sparse but well defined red giant branch and a HB populated in the red and variable regions; the turn--off luminosity suggests an age slightly lower than that of M~3, by  $\sim$1 Gyr \citep{vandenberg2000, catelan2001} or $\sim$2 Gyr \citep{stetson1999} --- see also \cite{hilker2006}.
	 \cite{catelan2001} investigated the CM diagram in the context of the ``second parameter'' problem, in comparison with that of M~3. By means of HB simulations, they found that a mass dispersion $\sigma$ $\sim$ 0.02 $M_\odot$ was required to reproduce the HB of M~3, while the HB of Pal~3 was consistent with a null mass dispersion.
\begin{figure}
\begin{center}
\includegraphics[width=8cm]{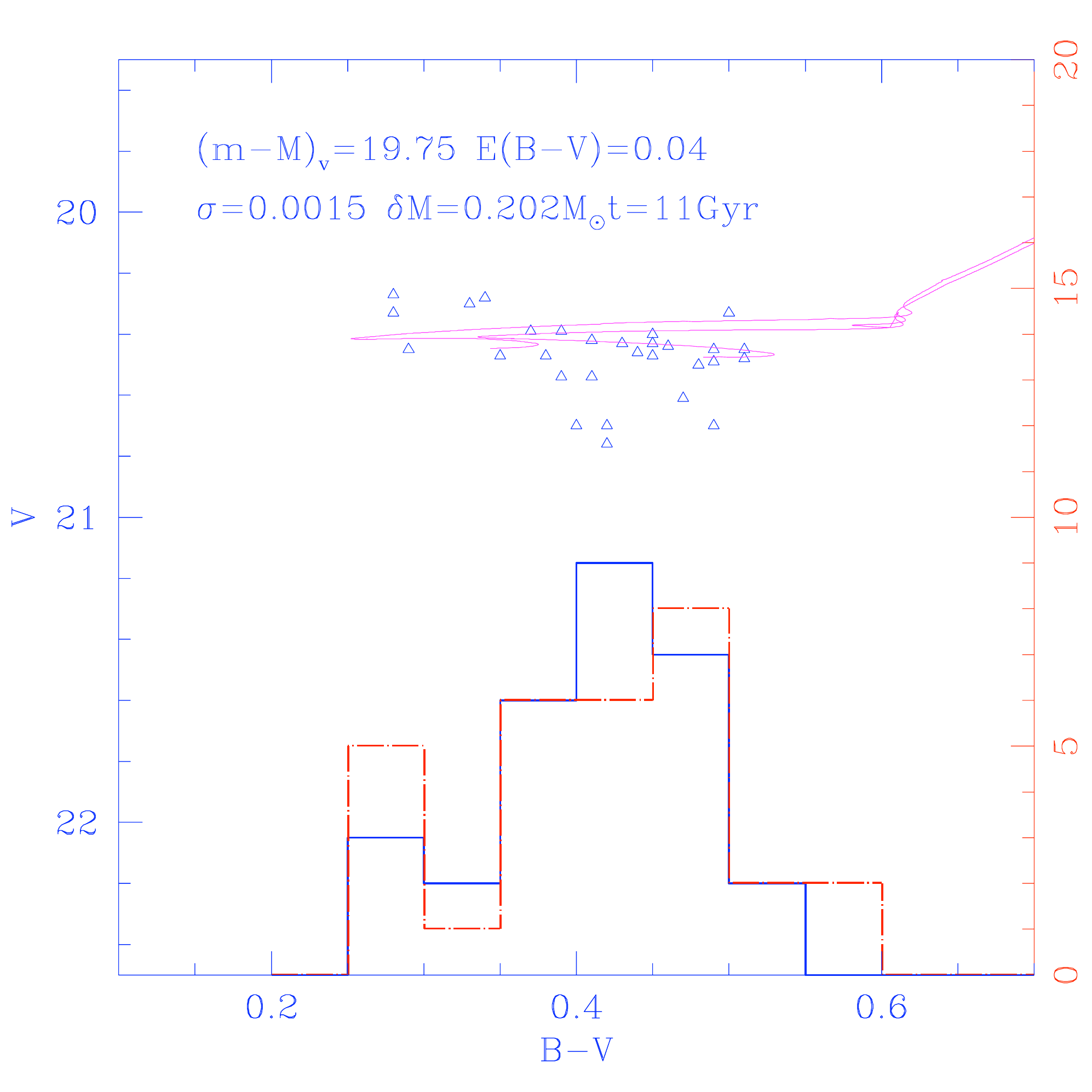}
\caption{HB stellar distribution as observed by Hilker , and superimposed simulation. 
}
\label{f2}
\end{center}
\end{figure}
      The chemical composition has been investigated by \cite{koch2009}; they obtained high resolution spectra for four red giants and determined the abundances for 25 elements ($\alpha$--, iron--peak, neutron--capture elements). The sample is limited, but a few results appear relatively safe: the $\alpha$--enhancement is compatible with that found in halo field stars and typical GCs as M~13; so are the Fe--peak and neutron--capture elements ratios. In addition, \cite{koch2010} find that the n--capture elements appear to derive from the {\it r}--process only, as observed only in the very metal poor field stars \citep{honda2007} and in the GC M~15 \citep{sneden2000}. Then, the n--capture patterns in Pal~3 do not require enrichment processes other than occurring in SN~II explosions. Besides, a Na--O anti--correlation is not evident, although a weak one cannot be ruled out by Koch et al's data.

     On the basis of the chemical and structural characteristics of the cluster, that is, absence of {\it s}--process elements and anti--correlations (admittedly, an absence not yet safely established)  we may consider Pal~3 as a good candidate to a FG--only cluster. For what concerns the dynamical evolution, let us note however that \cite{sohn2003} found a weak evidence of tidal extensions around the cluster, out to $\sim$ 4  times the tidal (truncation) radius. In view of the long relaxation times at the center and at $r_{h}$ (7 and 8 $10^{9}$ yr, respectively) and the estimated extremely long destruction time, we think that, in any case, such extra-tidal objects should constitute a minor side effect.
     
 In Fig.~\ref{f2} we show the result of a simulation of the HB distribution. The HB stars are taken from    \cite{hilker2006}. We find, as expected, a very small mass spread, $\sigma$=0.0015\msun, consistent with previous estimates \citep{catelan2001}. Therefore, the `HB criterion" corresponds well to the  other properties of this cluster, that is confirmed to be a FG--only cluster.

\begin{figure*}
\begin{center}
\includegraphics[width=8cm]{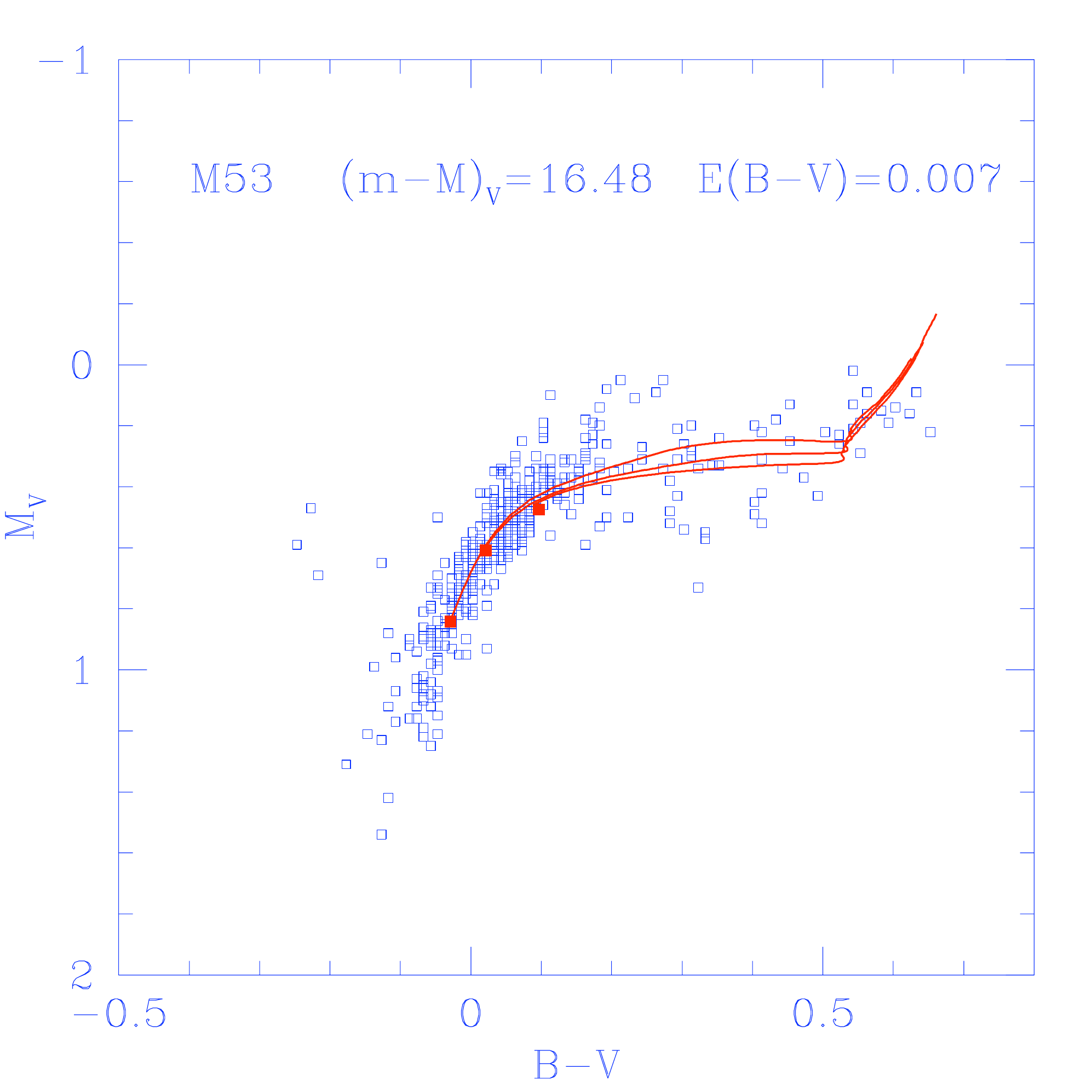}
\includegraphics[width=8cm]{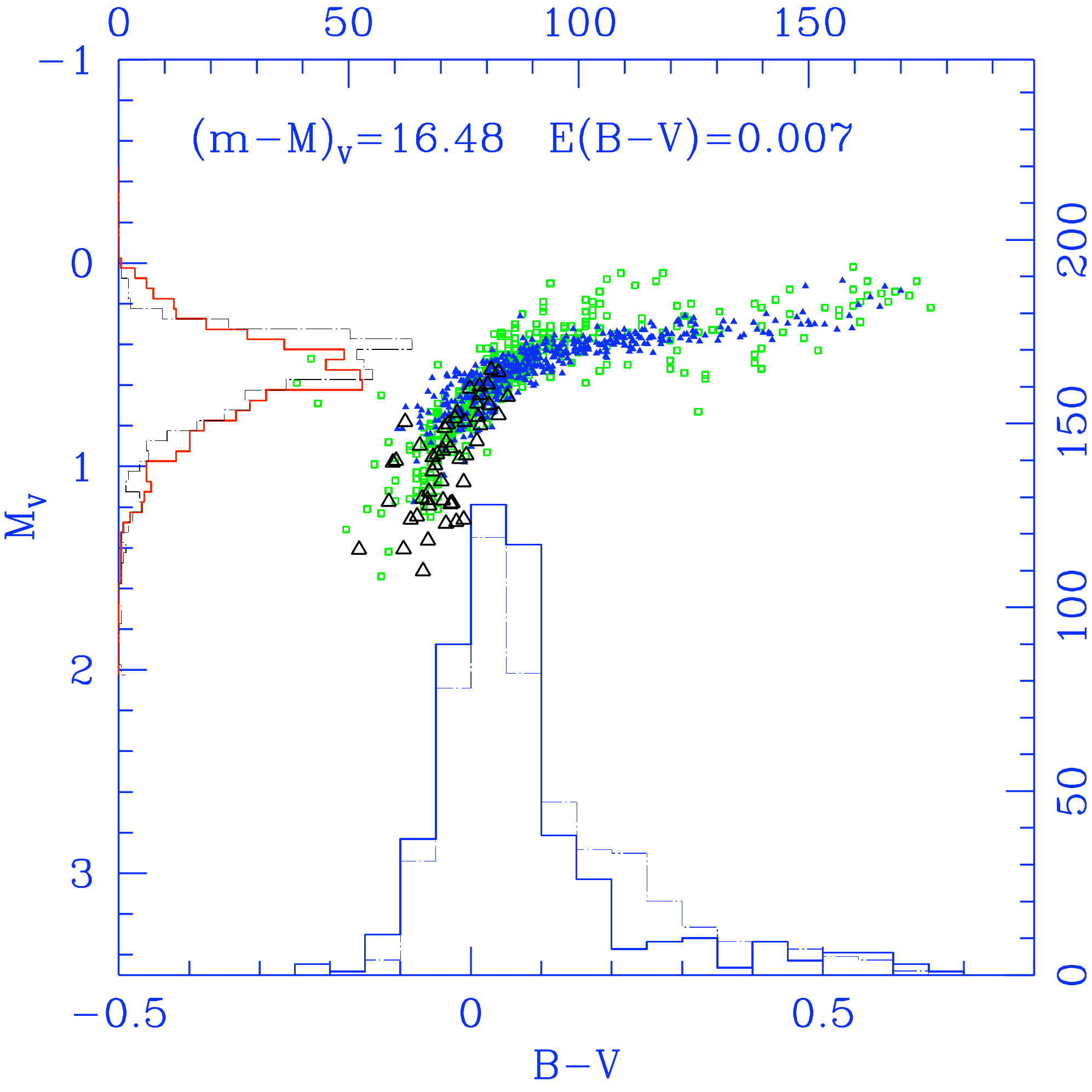}
\caption{Left panel: we show the HB tracks of M=0.68, 0.70 and 0.72\msun\ for Y=0.24, Z=0.0002 superimposed on the data by Rey et al. 2004. Right panel: a successful simulation (triangles) is superimposed on the data by Rey et al.
The blue filled triangles correspond to stars with Y=0.24, while the black open triangles are the stars with Y=0.26 and Y=0.27 (see text). The full histograms on the sides represent the number vs. magnitude and number vs. colour distributions, as observed (full lines) and simulated (dot--dashed lines).
}
\label{f3}
\end{center}
\end{figure*}

  \subsection{M~53 (NGC 5024)}
     This cluster is rather massive, with $M_{V}$=--8.70 and M=5$\times 10^5$\msun. The relaxation times at the center and at  $r_{h}$ are 5.8 $\times 10^{8}$ and 4.6 $\times 10^{9}$, respectively. Its destruction  time is given by \cite{gnedin1997} as about 30 Hubble times. It is rather far away, at 18.3 kpc from the Galactic center at a height above the Galactic plane of 17.5 kpc. It is generally considered among the very metal poor clusters ([Fe/H] = --1.99, Harris 2003).  The present \rh/\rj\ is 0.039, and its maximum value at the perigalactic distance of 15.5kpc \citep{allen2006} is only slightly larger (0.044). Therefore this cluster appears to have evolved always well inside its gravitational well.

     The first indication that M~53 may be an FG cluster\footnote{\cite{dantonacaloi2008} examined the HB structure of this cluster, and considered that it could be one of those GCs in which the first	 generation had been completely lost, together with NGC~6397 and M13.  For M~53 the main feature leading to this conjecture was the possible presence of a high nitrogen content in the integrated spectrum	 \citep{liburstein2003}. Here we have more information, that make us prefer the first stellar generation as the only one present, and not the second.} comes from the CM diagram. Its short HB barely reaches  $B - V$ = --0.05, with very few stars beyond this colour. In the cluster there are 59 known RR~Lyr variables of Oo type II \citep{kopacki2000};  a complete sample of the HB gives 12 red HB, 35 variables and 257  blue HB, these last ones almost all concentrated in a clump at a colour close to the  blue edge of the variable region \citep{rey1998}. At the metallicity of M~53, the HB tracks beginning in the colour region of this clump (B--V$\sim$0.0) evolve directly towards the red  \citep[][see also the left side panel of Fig.~3]{sweigart-gross1976, dicriscienzo2011}, and an HB of this type is the result of the evolution of a well defined HB  mass,  with  a very small dispersion in mass loss. The RR~Lyr and the red HB are the tail of the evolution of this typical evolving HB mass. This hypothesis is confirmed by the  simulations described in the following. A very similar situation is found in the upper HB of NGC~2419, and is examined in extensive detail in  \cite{dicriscienzo2011}.

     The presence of chemical inhomogeneities has been investigated by \cite{martell2008},  who studied the absorption bands of CN and  CH. What they found is ``a broad but not strongly bimodal distribution of CN bandstrength''. In their Fig. 6 they compare it with the situation in  NGC 6752: while in the latter cluster the CN  distribution is neatly bimodal, in M~53 it appears as a single Gaussian with a ``hunch'' on the shoulder toward higher abundances (see their Fig. 3, but see also Smolinski et al. 2011). Unfortunately, we can not infer information on the existence of an anticorrelation C--N from these data, since metal--poor clusters do not show the CH--CN anticorrelation, even in presence of the C--N bimodality  \citep[see the results for M~15 by][]{cohen2005}.  Nevertheless, on the basis of the relatively small range in C and N abundance variation in M~53, Martell et al. observe that may be ``the polluting material was not processed through the full CNO cycle'', in which case the cluster should  not show the Na--O anti--correlation. 
 
In order to better define at what level M~53 is dominated by the FG, we performed simulations of the HB morphology of this cluster. As the HB population is strongly peaked at V$\sim$17.0~mag and B--V$\sim$0.0, and the HB evolution is strictly redward, there is no possibility of reproducing the sparse blueward extension unless we hypothize that 1) either the mass loss on the RGB is slightly asymmetric; 2) or the bluest HB members have a slightly larger helium content, so that they have a smaller progenitor mass. In this latter case, a small SG population would be present, characterized by this small increase in helium.

Our simulations are performed according to this second framework. We assumed V$=$17.4 mag as the luminosity separating the main body of  the HB from a short tail of stars with a possibly different origin. We considered a total of 450 HB stars with V$<$17.4 mag, plus 50 fainter stars  with V$>$17.4 (see below). This number has been obtained by scaling the data by \cite{rey1998} with respect to the total number of RR~Lyr variables with known periods (59).  As for the RR~Lyraes, we tried to reproduce their period distribution. 

It was possible to reproduce the entire HB at V$<$17.4 attributing to the cluster an age of 12~Gyr, a standard helium content  Y = 0.24, and an average mass loss on the red giant branch of 0.113 \msun, with a dispersion $\sigma$(M)=0.015\msun; a Gaussian error of 0.03 mag has been associated to both B and V magnitudes.	 Other choices of parameters can be done with equivalent success, e.g. a smaller mass dispersion can be associated to larger observational errors. For the fainter HB we assume a slightly   higher Y content, from 0.26 to 0.27. A successful simulation, superimposed to Rey et al. data, with the mentioned choices for the distance modulus and the reddening, is shown in the right panel of Fig.~\ref{f3}.

\begin{figure}
\begin{center}
\includegraphics[width=8cm]{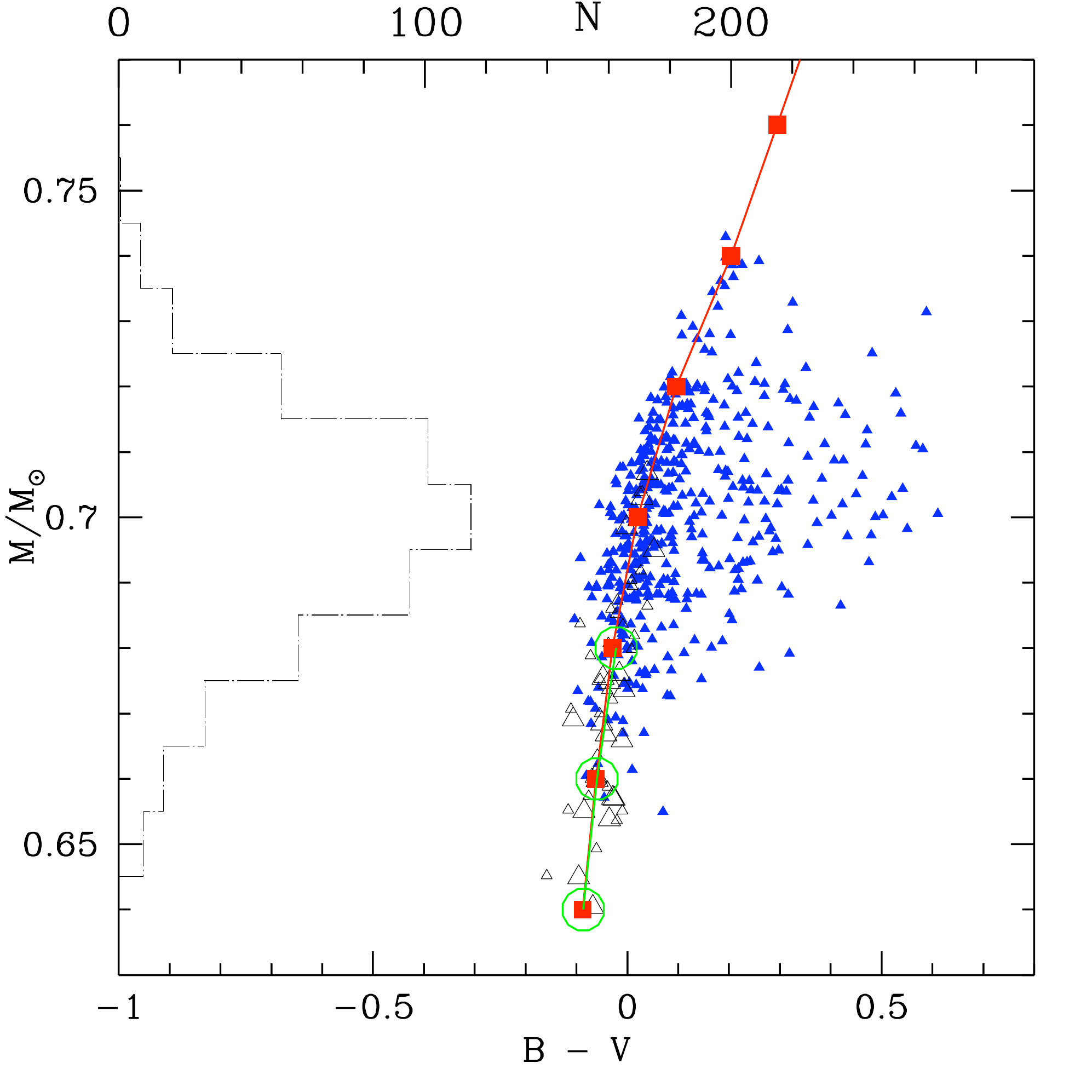}
\caption{HB mass distribution as a function of colour for the simulation of Fig.\ref{f3}. The blue filled triangles have Y=0.24, while the black smaller open triangles are the stars with Y=0.26 and the larger open triangles have Y=0.27. The histogram of number vs. stellar mass shown on the left is the sum of the Gaussian distribution of the 450 stars with Y=0.24 plus the slightly asymmetric extension towards smaller masses of the higher helium stars.
The dots mark the ZAHB of the models adopted for the simulation, having chemistry Z=0.0002 and Y=0.24, while the three open circles show the (coincident) ZAHB of the models with Y=0.28 that have been used to simulate the helium--increased population. Notice that no difference appears in the colours versus mass, but the Y=0.28 models are more luminous.
}
\label{f4}
\end{center}
\end{figure}

\begin{figure}
\begin{center}
\includegraphics[width=8cm]{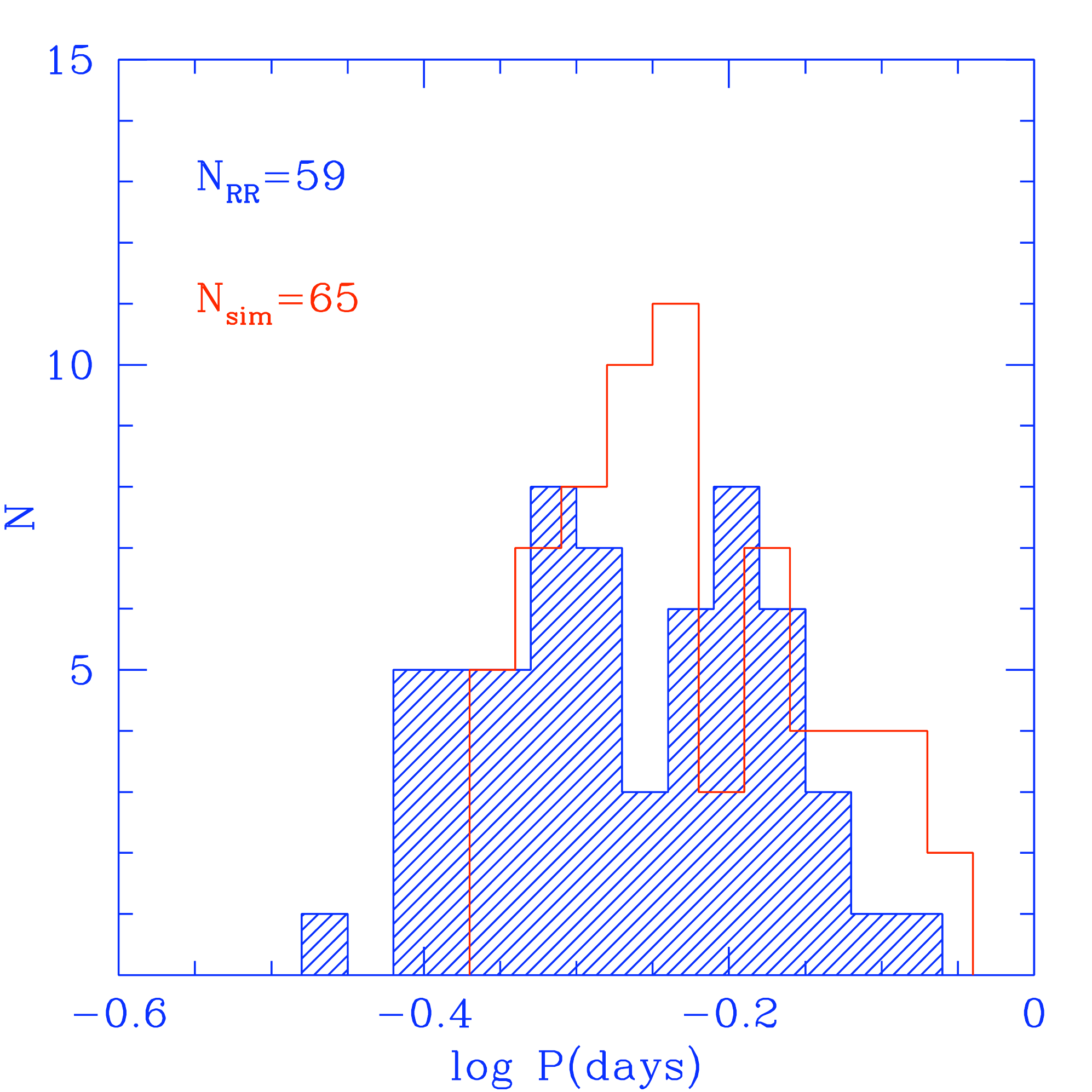}
\caption{RR~Lyr distribution (shaded) from Kopacki (2000) and Clement et  al. (2001), and a simulated distribution of periods, from the simulation of Fig.\ref{f3}.
}
\label{f5}
\end{center}
\end{figure}

\begin{figure}
\begin{center}
\includegraphics[width=8cm]{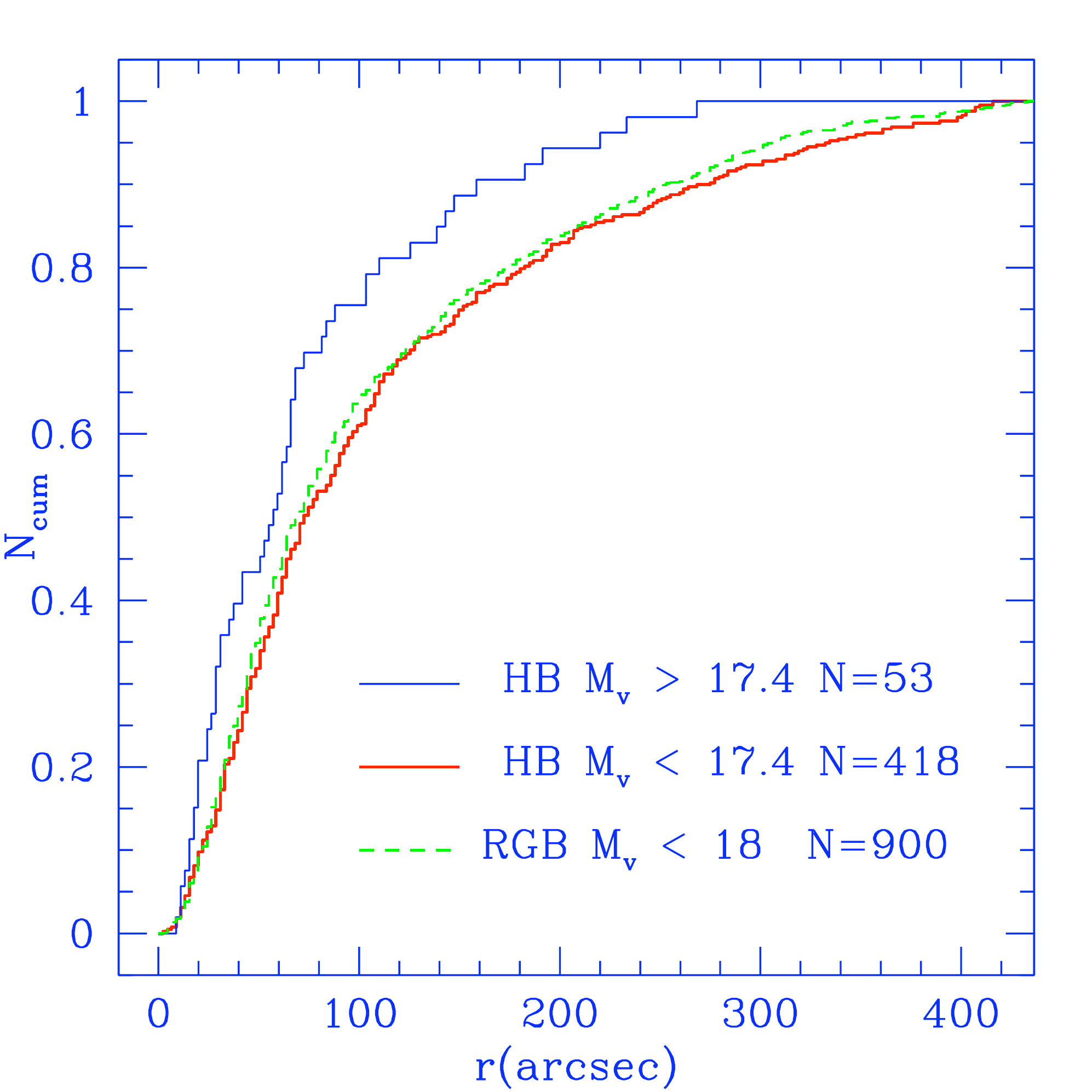}
\caption{Cumulative distribution of the majority of HB stars (thick line, red) is compared with the distributions of the bluest HB stars (upper line, blue) and of the red giants (dashed line, green). The sample has been divided by considering as ``extreme" all the HB stars at M$_v >$17.4mag.}
\label{f6}
\end{center}
\end{figure}
 
     The average evolving mass on the HB is 0.7\Msun. The simulated mass distribution as a function of the colour is shown in Fig.\ref{f4}, where we see that the variable and red HB stars represent the tail of  the distribution, and can easily be interpreted as evolving from the zero age HB at 0.69--0.72\msun. The zero age horizontal branch (ZAHB) of the models with Z=0.0002 and Y=0.24 is also shown on the simulation. We can appreciate that  very few ZAHB stars, with mass up to 0.74\msun, are present at colours  B--V$\sim$0.2, as discussed above.   
     As the RR~Lyr distribution with colour is dominated by stars already evolved from the ZAHB, it is statistically less constrained than the peak region. The total number and overall behaviour  of RR~Lyraes (see Fig.\ref{f5}) are reasonably reproduced by the  simulation presented in Fig.~\ref{f4}. We did not consider necessary to obtain a better agreement of the period distribution, even if in principle this appears possible, as we know by experience \citep{dantonacaloi2008}.  In our simulation, the slight asymmetry in the histogram of number versus mass shown in Fig.\ref{f4} is due to the presence of the small percentage of stars with higher helium content, but the same result could be obtained by assuming a slight asymmetry in the mass lost along the red giant branch.

We examined the relative spatial distribution of the faintest HB stars relative to the other HB members and to the red giants having V$<$18. The comparison employs \cite{rey1998} data, converting the pixel scale of their data base by knowing that 1 pixel = 0.22 arcsec and is shown in Fig.\ref{f6}.
While the red giant branch and the main body of HB stars have the same cumulative distribution,  the fainter HB stars appear indeed more concentrated. Is this an indication in favour of our interpretation, as the small SG would have formed in the very core of the cluster where the cooling flow concentrates the gas from the AGB ejecta? The study of dynamical mixing of two stellar generations, the second one formed exclusively in the core, is limited, until today, to the N--body simulations presented by \cite{dercole2008} in the context of their model for GC formation. Further study is certainly needed to understand whether the current properties of M~53 (Table 1) are consistent with the observed only partial mixing of SG and FG. M~53 is not unique in this respect. E.g. a central concentration for the SG is found  in NGC~3201 by \cite{kravtsov2010} and \cite{carretta2010b}, and in NGC~6752 by \cite{kravtsov2010}.
The second line of interpretation, that the giants in the cluster core suffer stronger dynamical interactions and are subject to a stronger mass loss, should also be carefully tested by modelling the interactions in the cluster core. At present, we suggest that spectroscopic  information (e.g. the presence or lack of high sodium and low oxygen in a small fraction of M~53 red giants) would be the best observationsl test of either hypothesis. At the present stage, we propose that current data show  some evidence that M~53 is a ``mainly--FG" cluster.

Visual inspection of their CM diagrams indicate that NGC 5634, NGC 5694 and NGC 6101 have HB characteristics very similar to those of M~53, so they can also be mainly--FG candidates, and it would be important to assess their chemical properties as well. Notice that only NGC~6101 is tidally filling: the other three clusters (including M~53) have a small ratio \rh/rj, and actually lie close to the line giving the position of a cluster of 10$^5$\msun\ with \rh=3~pc in the plane \rh/\rj\ {\it vs.} d$_{gc}$\ of Figure 2 in \cite{baumgardt2010}.

\section{NGC~2419 {\it versus} Pal~3 and M~53} 
 
Our analysis started from the consideration that the dynamical status of a cluster (tidally limited or not) at its formation time, or later on during the course of its life would determine its evolution and survival. However, since it is not straightforward  to go back from the present dynamical status to the previous dynamical evolution, we resorted to a pure photometric parameter (the morphology of the HB) to select FG--clusters. In this way we have not only recovered that Pal~3 is an FG--only cluster \citep{koch2009,koch2010}, but we have also shown that M~53 may be an example of a mainly--FG cluster. 

At this point it is relevant to enquire how would be like a cluster, in principle similar to M~53, in which the SG appears to be a relevant constituent. We consider NGC~2419, a metal poor, far away, isolated cluster (Table 1). It is twice as massive as M~53, and much farther from filling its Galactic tidal radius than M~53: the comparison between the truncation radius and Jacobi radius  provides a ratio   \rt/\rj~= 0.57 for M~53 and only 0.28 for NGC~2419. 
\begin{figure}
\begin{center}
\includegraphics[width=8cm]{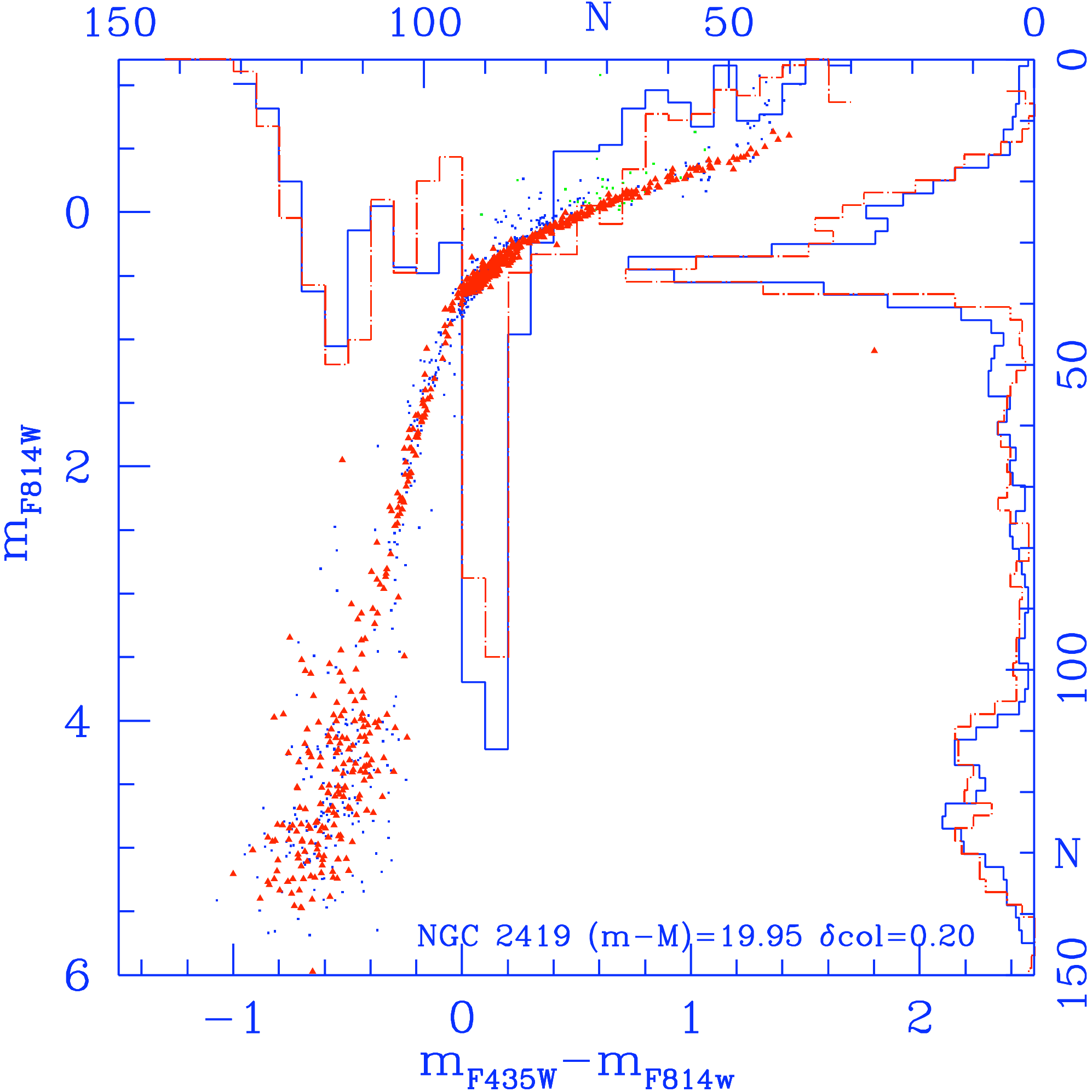}
\caption{We show the HB data for NGC~2419 in the Hubble Space Telescope near infrared magnitude  m$_{\rm F814W}$\ versus the color m$_{\rm F435W}$--m$_{\rm F814W}$\ presented in Di Criscienzo et al. (2011b). The histogram of colour and magnitude distributions are also shown as full (blue). Superimposed we show a simulation of the entire HB, obtained by fitting the luminous part with 390 stars having standard helium abundance Y=0.24, mass loss $\Delta$M=0.073\msun\ and spread in the HB masses $\sigma$=0.008, the middle part with 90 stars having the same Y=0.24 and $\Delta$M=0.22\msun, $\sigma$=0.05\msun; the blue hook is reproduced with 160 stars with Y=0.42, $\Delta$M=0.110\msun\ and $\sigma=0.01$\msun. Further assumption of the models for the blue hook stars are provided in Di Criscienzo et al. 2011b. The histograms of the simulation are dot--dashed. }
\label{f7}
\end{center}
\end{figure}
Given the similarity in heavy element abundance, we can compare the HBs of  M~53 and NGC~2419 (Fig.~\ref{f1}). As mentioned  before, the brighter regions of the HB are very similar: the same concentration of HB members  in a very small color interval on the blue side of the variable region, with RR~Lyr and red HB stars likely the product of these blue stars, as discussed in the simulation of M~53 HB (Fig.\ref{f3}). But the HB in NGC~2419 continues, with a lower star density, well beyond the limit in M~53, ending with the most populated ($\sim$30\% of the HB stars!) blue hook known in Galactic  GCs \citep{ripepi2007}. Although \cite{sandquist2008} argue that the symmetric distribution of stars along the MS  favours a single stellar population model for this cluster, \cite{dicriscienzo2011} have discussed that both the HB morphology and the colour distribution of the giants suggest the presence of two well separated populations, one of which has a very high helium content. They also show that the partial asymmetry that would result in the MS colour distribution would be hidden by the present photometric errors. If this is the case, the dynamical evolution of this cluster can not have occurred in isolation, as a much larger initial mass would have been required to provide the 30\% very helium rich population presently found in NGC~2419.
The small spread in calcium present among the giants in NGC~2419 \citep{cohen2010} also shows that this cluster was able to retain at least some SN~II ejecta, spectroscopically excluding the FG--only possibility. 
In Fig.~\ref{f7} we show a simulation of the whole HB of NGC~2419, well reproducing the whole extension in colour and magnitude of this extreme HB. In order to reproduce the upper HB we assumed Y=0.24 and $\sigma$=0.008\msun. The long tail, between the more luminous peak of stars and the blue hook, is reproduced by 90 stars, and assuming the same helium content and a broader and larger mass loss ($\Delta$M=0.22\msun\, $\sigma$= 0.05\msun). A similar fit is obtained by assuming stars with increased and variable helium content, and a smaller spread in mass loss, as described above for M~53.  For the blue hook we assume Y=0.42 and $\sigma$=0.01\msun. The blue hook simulation follows the prescriptions explained in \cite{dicriscienzo2011}, to which we refer for details. 

\section{Discussion}
We adopted a simple photometric criterion (the extension of the HB) to select candidates for FG clusters from existing large databases of CM diagrams of GCs. The list of 17 candidates is tentatively divided into 8 FG--only clusters (that could not form an SG at all) and 9  mainly--FG clusters (that could form an SG, but did not lose most of the FG mass). The small percentage of mainly--FG clusters indicates that very few of the clusters that develop the SG do not lose a high fraction of their initial mass, and we suggest that non isolated clusters are destroyed by the dynamical interaction with the Galactic tidal field, unless they are able to form a SG, whose dynamical mixing with the FG stars allows the cluster to survive \citep[see also][]{dvmessenger2008}.

We studied in more detail the HBs of  Pal~3,  M~53 and, for comparison,  NGC~2419. We found a similarity between  the HB in M~53 and the upper HB in NGC~2419, the total cluster mass being the only  evident difference between them. Given their structural similarity, we may  be witnessing the influence of total mass {\it only} on the first  evolutionary stages in GC life (or at least, in the very metal poor ones).  The three clusters represent very different  evolutionary cases:\par 
1) Pal~3 is consistent with hosting a FG--only population. \par 2) M~53, a much more massive cluster, has a HB consistent with an almost pure FG population; by examining synthetic models for its HB stellar distribution, we suggest that a small second generation may be present (mainly--FG cluster), given by the bluest and faintest HB stars, mostly concentrated in the cluster inner regions.  \par 3) NGC~2419, more than double the mass of M~53 and close in mass to the  most massive clusters in the Galaxy, with an apparently ``evident'' chemical and  dynamical isolated evolution, exhibits a consistent blue hook, a crucial signature of the presence of multiple star generations.  So it is reasonable to expect strong chemical anomalies in this cluster (O--Na and Mg--Al anticorrelations) --  not yet observed, given its extreme distance. The presence of a small spread in calcium \citep{cohen2010} testifies however that this cluster was able to retain at least some SN~II ejecta, spectroscopically excluding the FG--only possibility. 

  We have presented reasons to support the hypothesis that M~53 is a mainly--FG cluster. Were this the case, and if, as likely,  the chemical and dynamical evolution of the cluster had taken place within its tidal radius, the interesting possibility would arise of estimating the percentage of second generation stars resulting from such evolution, after estimating the stellar losses due to two--body interactions during the Galactic lifetime. 
To  clarify this scenario, it would be important to obtain a CM diagram of M~53  with modern  telescopes, in order to establish with higher precision the colour  of the population peak on the HB. In fact, the large part of the HB members  with smaller colours can be assumed to belong to the second generation,   once excluded the possibility of an asymmetry in mass loss along the RG  branch. This derives from the fact that the evolution of ZAHB members bluer  than the RR~Lyrae strip develops strictly redward. At present, the fraction of the SG seems to be $\sim$ 0.1 (=50/500), but this is surely an approximate value.

  The consistency of HB morphologies with chemical characteristics  known at present, even if encouraging, must  be substantiated by spectroscopic investigations of the giant branches of  all the three clusters, to put in evidence the presence/absence of, e.g.,  the Na--O anticorrelation. We hope that this information will help to  establish a few firm points in this complex subject.  
 
 \section{Acknowledgments} 
This work has been supported through PRIN INAF 2009 "Formation and Early
Evolution of Massive Star Cluster". We thank Enrico Vesperini for useful discussion and comments, and M. Hilker and P. Stetson for information concerning their data for Pal~3.  The anonymous referee is thanked for his/her stimulating report.

\label{lastpage}

\end{document}